\documentclass[12pt]{article}
\usepackage[dvips]{graphicx}
\let\jnfont=\rm
\def\NPB#1,{{\jnfont Nucl.\ Phys.\ }{\bf B#1},}
\def\PLB#1,{{\jnfont Phys.\ Lett.\ B }{\bf #1},}
\def\PRD#1,{{\jnfont Phys.\ Rev.\ D }{\bf #1},}
\def\PRL#1,{{\jnfont Phys.\ Rev.\ Lett.\ }{\bf #1},}
\def\ZPC#1,{{\jnfont Z.~Phys.\ C }{\bf #1},}
\def\ETslash{\not{\hbox{\kern-4pt $E_T$}}}
\newcommand{\gsim}{\mathrel{\lower4pt\hbox{$\sim$}}
\hskip-12.5pt\raise1.6pt\hbox{$>$}\;}
\newcommand{\lsim}{\mathrel{\lower4pt\hbox{$\sim$}}
\hskip-12.5pt\raise1.6pt\hbox{$<$}\;}
\let\to=\rightarrow
\begin{document}

\baselineskip=0.25in
\begin{flushright}
AMES-HET-00-06 \\
July 2000 
\end{flushright}
\vspace{0.2in}
\begin{center}
{\Large Probing R-violating top quark decays at hadron colliders}
\vspace{.3in}

 K.J. Abraham$^a$, Kerry Whisnant$^a$, Jin Min Yang$^b$, 
 Bing-Lin Young$^{a,b}$ 
\vspace{.3in}

{\small \it
$^a$ Department of Physics and Astronomy, Iowa State University,
     Ames, Iowa 50011, USA\\
$^b$ Institute of Theoretical Physics, Academia Sinica, Beijing 100080, China}
\end{center}
\vspace{.5in}

     \begin{center} ABSTRACT  \end{center}
We examine the possibility of observing exotic top quark decays
via R-violating SUSY interactions at the Fermilab Tevatron and CERN LHC. 
We present cross-sections for $t\bar t$ production followed by the 
subsequent decay of either $t$ or $\bar t$
via  the R-violating interaction while the other undergoes the 
SM decay.   With suitable kinematic cuts, we find that 
the exotic decays can possibly be detected over standard model
backgrounds at the future runs of the 
Tevatron and LHC, 
but not at Run 1 of the Tevatron due to limited statistics.
Discovery limits for R-Violating couplings in the top
sector are presented. 


\newpage
\section{Introduction}
\label{sec1}
The top quark, with a mass of the order of the electroweak symmetry breaking 
scale, is naturally considered to be related to new physics. 
Run~1 of the Fermilab Tevatron has small statistics on top quark events
and thus leaves plenty of room for new physics to be discovered at
the upgraded Tevatron~\cite{tev2000} in the near future. 
Due to higher statistics, the $t\bar t$ events at the upgraded Tevatron
are expected to provide sensitive probes for new physics~\cite{tt1}.  
The most popular model for new physics is the Minimal Supersymmetric 
Model (MSSM)~\cite{mssm}. In this model, $R$-parity~\cite{rp}, 
defined by $R=(-1)^{2S+3B+L}$ with spin $S$, baryon-number $B$ and  
lepton-number $L$,  is often imposed on the Lagrangian to maintain 
the separate conservation of $B$ and $L$.  As a consequence the 
sparticle number is conserved. 
Since instanton effects 
induce miniscule violations of baryon and 
lepton number 
\cite{tHooft}, $R$-parity conservation is not dictated by any known
fundamental principle such as gauge invariance or renormalizability. 
If $R$-parity is strictly conserved, it is conceivable that the 
conservation comes from some hitherto unidentified fundamental principle.
Hence $R$-parity violation should be vigorously searched for. 

The most general superpotential of the MSSM, consistent with the
$SU(3)\times SU(2)\times U(1)$ symmetry, supersymmetry, and
renomalizability 
also contains $R$-violating interactions which are given by 
\begin{eqnarray}\label{WR}
{\cal W}_{\not \! R}=\frac{1}{2}\lambda_{ijk}L_iL_jE_k^c
+\lambda_{ijk}^{\prime} \delta^{\alpha\beta} L_iQ_{j\alpha}D_{k\beta}^c
+\frac{1}{2}\lambda_{ijk}^{\prime\prime}\epsilon^{\alpha\beta\gamma}
U_{i\alpha}^cD_{j\beta}^cD_{k\gamma}^c
+\mu_iL_iH_2,
\end{eqnarray}
where $L_i(Q_i)$ and $E_i(U_i,D_i)$ are the left-handed
lepton (quark) doublet and right-handed lepton (quark) singlet chiral 
superfields. The indicies $i,j,k$ are generation indices, 
$\alpha$, $\beta$ and $\gamma$ are the color indices, 
$c$ denotes charge conjugation, 
and $\epsilon^{\alpha\beta\gamma}$ is the total antisymmetric tensor
in three-dimension. $H_{1,2}$ are the Higgs-doublets chiral superfields.
The coefficients $\lambda$ and $\lambda^{\prime}$ are the coupling
strengths of the $L$-violating
interactions and $\lambda^{\prime\prime}$ those of the $B$-violating
interactions.  The lower bound of the proton lifetime imposes 
very strong conditions on the simultaneous presence of both $L$-violating 
and $B$-violating interactions~\cite{proton} and hence the strength of
the couplings.  However, the existence 
of either $L$-violating or $B$-violating couplings, but not both at the
same time, does not induce nucleon decays and therefore the $R$-parity
violating couplings are less constrained. This separate $L$ and $B$
violation 
is usually assumed in phenomenological analyses.

The study of the phenomenology of R-violating supersymmetry was
started many years ago~\cite{rvmssm}.  
Some constraints on the $R$-parity violating couplings have been obtained 
from various analyses,  such as perturbative unitarity~\cite{pert},
$n-\bar n$ oscillation~\cite{nn}, $\nu_e$-Majorana mass~\cite{nue-mass}, 
neutrino-less double $\beta$ decay~\cite{d-beta}, charged current 
universality~\cite{Barger}, $e-\mu-\tau$ universality~\cite{Barger},
$\nu_{\mu}-e$ scattering~\cite{Barger}, atomic parity violation~\cite{Barger}, 
$\nu_{\mu}$ deep-inelastic scattering~\cite{Barger}, 
$\mu-e$ conversion~\cite{mu-e}, $K$-decay~\cite{Bdecay},
$\tau$-decay~\cite{tau}, $D$-decay~\cite{tau}, $B$-decay~\cite{Bdecay}
and $Z$-decay at LEP I~\cite{Zdecay}. 
As reviewed in Ref.~\cite{review}, although many such couplings have been
severely constrained, the bounds on the top quark couplings 
are generally quite weak. 
This is the motivation for the phenomenological study of R-violation in 
processes involving the top quark.

The production mechanisms of top pairs and single top 
in $R$-violating SUSY at the upgraded Tevatron have been examined
in~\cite{hadron1} and~\cite{hadron2}, respectively. 
In addition, the $R$-violating couplings can induce exotic 
decays for top quark at an observable level.
For example, the top quark FCNC decays induced by $R$-violating couplings 
~\cite{tcv1} can be significantly larger than those in the MSSM with
R-parity conservation~\cite{tcv2}. 
If we allow the co-existence of two $\lambda'$ couplings, we have the
new decay modes, such as 
$t\to \ell  \tilde d \to  \ell^+ \ell^- u$~\cite{atwood}.
The bilinear term $\mu_iL_iH_2$ can also induce some new decays 
for the top quark, as studied in~\cite{campos}. 

In this work, we focus on the explicit trilinear couplings and 
assume only one trilinear coupling exists at one time. Then the 
possible exotic top decay modes are 
\begin{eqnarray}\label{BV}
t\to \tilde{\bar d_i}  \bar d_j,~\tilde{\bar{d}}_j \bar{d}_i
          \to \bar d_i  \bar d_j \tilde \chi^0_1 
\end{eqnarray}
induced by the $B$-violating $\lambda''_{3ij}$, and
\begin{eqnarray}\label{LV}
t\to e^+_i  \tilde d_j ~ \tilde{e}_i d_j
                 \to e^+_i d_j \tilde \chi^0_1 
\end{eqnarray}
induced by the $L$-violating $\lambda'_{i3j}$. Here the subscripts
$i,j$ are family indices and $ \tilde \chi^0_1$ is
the lightest neutralino which, in our analysis, is assumed to be the lightest 
super particle (LSP)  as favored in the MSSM 
where the SUSY breaking is propagated to the matter sector by 
gravity\footnote{If the SUSY breaking is mediated 
by gauge interactions, the LSP is expected to be the gravitino.}.  
The sfermions involved in these decays can be on-shell or virtual,
depending on the masses of the particles involved.

Among the exotic decays in (\ref{BV}) and (\ref{LV}), the relatively 
easy-to-detect modes are those induced by $\lambda''_{33j}$ 
($j=1,2$)\footnote{$\lambda''_{333}$ does not exist since $\lambda''_{ijk}$ is 
antisymmetric in the last two indices.}
and $\lambda'_{i33}$ ($i=1,2,3$) because their final states contain a 
$b$-quark which can be tagged.  One of the $L$-violating channels,
i.e., $t\to \tilde \tau b ~({\rm or~} \tau \tilde b)$
induced by $\lambda'_{333}$ has been studied in~\cite{magro}.
 So in our analysis we focus on the cases of $\lambda'_{133}$ and 
$\lambda'_{233}$ for $L$-violating couplings, and $\lambda''_{331}$ 
and $\lambda''_{332}$ for $B$-violating couplings.
Since the decay induced by  $\lambda'_{133}$ has the similar final
states to that induced by $\lambda'_{233}$, we take the presence of
$\lambda'_{233}$ as an example. For the same reason, we take
the presence of $\lambda''_{331}$ as an example in $B$-violating case.  
The Feynman diagrams for these two decays induced by 
$\lambda''_{331}$ and  $\lambda'_{233}$ are shown in Figs. 1 and 
2, respectively.

In our analysis we consider $t\bar t$ events where one ($t$ or $\bar t$) 
decays via  R-violating coupling while the other decays by the SM 
interaction.  The SM decay will serve as the tag of the $\bar{t} t$
event.  Furthermore, the penalty of the suppressed $R$-violation 
coupling is paid only once.  Top spin correlations are
taken into account in our calculation. 

Note that the LSP ($ \tilde \chi^0_1$) is no longer stable when R-parity is 
violated. In case just one R-violating top quark coupling does not vanish,
the lifetime of the LSP will be very long, depending the coupling and the
masses of squarks involved in the LSP decay chain (cf. the last paper 
of~\cite{review}). Thus it is generally assumes that the LSP decays outside 
the detector~\cite{outside}. We will make this assumption in our analysis.  
This paper is orgnized as follows.  
In Sec.~\ref{sec2}, we investigate the potential of observing the 
$B$-violating top quark decay at the Tevatron and LHC, and present 
numerical results.
In Sec.~\ref{sec3} we present similar results for $L$-violating decay. 
Finally in Sec.~\ref{sec4} we present a summary and discussion.

\section{Searching for $B$-violating decay}
\label{sec2}

\subsection{Signal and background}

To probe the decay $t\to \bar b \bar d  \tilde \chi^0_1$ in Fig. 1,
we consider the final states given by $t\bar t$ production
where one (say $t$) decays via  the coupling $\lambda''_{331}$
while the other (say $\bar{t}$ has the SM decays to serve as the
tag of the $\bar{t} t$ event. 
Due to the large QCD background at hadron colliders, we do not
search for the all-jets channel despite
of its higher rate. Instead, we search for the signal given by
$t\bar t$ events followed by $t\to \bar b \bar d  \tilde \chi^0_1$ and
$\bar t \to W^- \bar b \to \ell \bar \nu \bar b$ ($\ell=e,\mu$).
Then the signature is a lepton, three jets containing two b-jets or 
two $\bar b$-jets, and missing energy ($\ell+3j/2b +\ETslash$).
We require that two $b$-jets are tagged in the signal. The efficiency
for double b-tagging is assumed to be $42\%$~\cite{tev2000}.

Note that the present events have the unique signal of the two same
sign b-quarks. In our analysis, to be conservative, we assume that 
the tagging can not distinguish a b-quark jet from $\bar b$-quark jet. 
Then the SM backgrounds are mainly from 
\begin{itemize}
\item[{\rm(1)}]  $t\bar t\to W^-W^+b\bar b$  followed by
                 $W^-\to \ell \bar \nu$ ($\ell=e$,$\mu$) and 
                 $W^+\to \tau^+ \nu$
                  with the $\tau$ decaying into a jet plus
                  a neutrino;
\item[{\rm(2)}] $t\bar t\to W^-W^+b\bar b$ followed by
                $W^-\to \ell \nu$  ($\ell=e$,$\mu$) and  $W^+\to q \bar q'$.
                This process  contains an extra quark jet 
                and can only mimic our signal if the quark misses 
                detection by going into the beam pipe. We assume 
                this can only happen when the light quark jet has 
                the pseudo-rapidity greater than about 3 or the transverse 
                momentum less than about 10~GeV.  
\item[{\rm(3)}] $Wb\bar b j$ which includes single top quark production 
                via the quark-gluon process $qg\rightarrow q't\bar b$
                as well as non-top processes~\cite{boos}. 
\end{itemize} 

\subsection{Numerical calculation and results} 
We calculated the signal and background cross sections with the 
CTEQ5L structure functions~\cite{cteq5L}. We assume $M_t=175$~GeV
and take $\sqrt s=2$~TeV for the upgraded Tevatron and $\sqrt s=14$~TeV for
the LHC. 

As shown in Fig. 1, there are two contributing graphs. 
Since among the down-type squarks the sbottom is most likely to be lighter 
than other squarks (we will elaborate on this later), 
we assume the first graph in Fig.1 gives the dominant contribution.
(If the $\tilde d$ is as light as the sbottom, the second diagram 
in Fig.1 has to be taken into account. Then our results for the signal 
rate should be quadrupled. To be conservative, we do not consider this case.)

For the total width of the sbottom involved in our calculation, 
we note that since only a light sbottom is meaningful to our analysis 
(as will be shown in our results), its dominant (or maybe the only) 
decay mode is $\tilde b\to b \tilde \chi^0_1$. The charged current 
decay mode $\tilde b\to t \tilde \chi^+_1$ is kinematically forbidden for
a light sbottom in our analysis. We do not consider the strong decay mode
$\tilde b\to b \tilde g$ since the gluino $\tilde g$ is likely
to be heavy~\cite{CDF}. 

The signal cross section is proportional to $|\lambda''_{331}|^2$. 
We will present the signal results normalized to  $|\lambda''_{331}|^2$.
 The  signal cross section is very sensitive to the sbottom
mass. We will vary it to see how heavy it can be for the signal to
be observable. Other SUSY parameters involved are the lightest neutralino 
mass and its coupling to sbottom, which are determined by the parameters
$M, M^{\prime},\mu$ and $\tan\beta$.
$M$ is the $SU(2)$ gaugino mass and $M^{\prime}$ is the hypercharge
$U(1)$ gaugino mass. 
$\mu$ is the Higgs mixing term ($\mu H_1 H_2$) in the superpotential.
$\tan\beta=v_2/v_1$ is the ratio of the vacuum expectation values of the 
two Higgs doublets. We work in the framework of the general MSSM. But we 
assume the grand unification of the gaugino masses, which gives the relation
 $M^{\prime}=\frac{5}{3}M\tan^2\theta_W\simeq 0.5 M$.
Then for the three independent parameters $M, \mu$ and $\tan\beta$,
 we choose a representative set of values 
\begin{eqnarray} \label{para}
M=100 {\rm ~GeV}, \mu=-200 {\rm ~GeV}, \tan\beta=1. 
\end{eqnarray}
They yield the lightest chargino and neutralino masses as 
$m_{\tilde\chi^+_1}=120$~GeV ,  $m_{\tilde\chi^0_1}=55$~GeV.
Thus this set of values are allowed by the current experimental bounds  
on the chargino and neutralino masses, which are about 90~GeV and 45~GeV,
respectively~\cite{LEP}.  

We simulate detector effects
by assuming Gaussian smearing of the energy of the charged final state
particles, given by:
\begin{eqnarray}
\Delta E / E & = & 30 \% / \sqrt{E} \oplus 1 \% \rm{,~for~leptons~,}
\label{eq:smear1}\\
             & = & 80 \% / \sqrt{E} \oplus 5 \% \rm{,~for~hadrons~,}
\label{eq:smear2}
\end{eqnarray}
where $\oplus$ indicates that the energy dependent and independent
terms are added in quadrature and $E$ is in GeV.

The basic selection cuts are chosen as
\begin{eqnarray} \label{basic1}
p_T^{\ell}, ~p_T^{jet},~
p_T^{\rm miss}&\ge& 20 \rm{~GeV} ~, \\ 
\eta_{jet},~\eta_{\ell} &\le& 2.5 ~, \\
\label{basic3}
\Delta R_{jj},~\Delta R_{j\ell} &\ge& 0.5 ~.
\end{eqnarray}
Here $p_{T}$ denotes transverse momentum, $\eta$ is the pseudo-rapidity,
 and $\Delta R$ is the separation in the azimuthal angle-pseudo rapidity 
plane $(~\Delta R= \sqrt{(\Delta \phi)^2 + (\Delta \eta)^2}~ )$ between 
a jet and a lepton or between two jets.

We notice that for the  background process (2) and (3)
the missing energy comes only from the neutrino of the W decay, while for the 
signal events the missing energy contains an extra neutralino.
From the transverse momentum of the lepton $\vec P_T^{\ell}$ and the missing 
transverse momentum $\vec P_T^{\rm miss}$,  
we construct the transverse mass as 
\begin{equation}\label{m_T}
m_T(\ell,p_T^{\rm miss}) = \sqrt{ (|\vec P_T^{\ell}|+|\vec P_T^{\rm miss}|)^2
- (\vec P_T^{\ell}+\vec P_T^{\rm miss})^2}.
\end{equation}
As is well-known, if the two components, i.e., $\ell$ and $p_T^{\rm miss}$
in our case, are from the decay of a parent particle, the transverse mass
is bound by the mass of the parent particle.  So for $Wb\bar b j$ background 
events $m_T(\ell,p_T^{\rm miss})$ is always less than $M_W$ and peaks
just below $M_W$.  However, kinematic smearings can push the bound and the
peak above $M_W$. 
In order to substantially suppress the large backgrounds (2) and (3)
we apply the following cut
\begin{equation} 
m_{T}(\ell,p_T^{\rm miss}) > 120 {\rm ~GeV}.
\end{equation}
We found that the above strong $m_T(\ell,p_T^{\rm miss})$ cut suppresses
the background process (2) and (3) by roughly three orders of magnitude
for the smearing in Eqs.~(\ref{eq:smear1}) and (\ref{eq:smear2}), so
that they are much smaller than the other backgrounds we are
considering. But since background process (1) contains three neutrinos
from different parent particles, it is not supressed by the
$m_T(\ell,p_T^{\rm miss})$ cut to a negligible level. There is some
model dependence involved the treatment of $\tau$ hadronization. To
avoid having to consider each of the many hadronic decay modes
separately, we assume the invariant mass of the 
outgoing hadrons to be distributed uniformly from  $m_{\pi}$ to $m_{\tau}$.
Furthermore, we assume a uniform angular distribution in the phase allowed
by the invariant mass of the outgoing jet. This assumption is probably 
reasonable in light of the fact that the parent $\tau$ is heavily boosted 
in the lab frame.

With the above selection cuts, the signal and background cross sections
are given in Table 1. We see that the signal-to-background ratio can
be quite large for light sbottom mass ($\lsim 160$~GeV), in which the
intermediate sbottom can be materialized as a real particle. When the sbottom
becomes heavier than the top quark and thus can only appear as a
virtual state, the cross section is severely suppressed by the small
branching ratio of the decay.

From the results for Tevatron (1.8 TeV) in Table 1 we conclude 
that the luminosity Run 1 (0.1 fb$^{-1}$) is too low to detect such decays.
However, due to the much larger statistics of Run 2 (2 fb$^{-1}$) 
and Run 3 (30 fb$^{-1}$), it is possible to observe such decays in these 
coming runs of the Tevatron.  Using the discovery criteria 
$ S \ge 5 \sqrt{B}$,  the discovery limits of $\lambda''_{33j}$ versus 
the sbottom mass at Run 2, Run 3 (30 fb$^{-1}$) and LHC (10 fb$^{-1}$) are 
plotted in Fig. 3. The region above each curve is the corresponding region 
of discovery. 
Since the current bounds on  $\lambda''_{331}$ from the LEP I Z-decay
are of ${\cal O}(1)$ for sfermion mass heavier than 100~GeV
~\cite{Zdecay},  we see that for a light sbottom, we have a good chance to  
observe such decays if  $\lambda''_{331}$ is not far below its current 
upper bounds. In case of nonobservation, meaningful bounds at 95\% C.L.
can be set, as shown in Fig.4.   

Our results for $\lambda''_{331}$ can be applied to the case of    
$\lambda''_{332}$. Since the current bound from $Z$-decay is the same  
on both couplings~\cite{Zdecay}, our conclusions on
$\lambda''_{331}$ can be applied to  the case of  $\lambda''_{332}$.    

\section{ Searching for $L$-violating decay}
\label{sec3}

For  $L$-violating decay $t\to  \mu^+  b   \tilde \chi^0_1$,
there are two contributing graphs, as shown  in Fig.2.
The first graph proceeds through exchanging a sbottom while
the second through exchanging a slepton. As in Sec.~\ref{sec2},
we assume sbottom can be light and thus concentrate on the first
graph. In the opposite case that the slepton is light and sbottom is
heavy, our following results still hold with the replacement of 
sbottom mass by slepton mass. If both sbottom and slepton are light
and approximately degenerate (which is quite unlikely in the supergravity 
scenario of supersymmetry breaking, as will be elaborated on later), 
then our results for the signal rate should be quadrupled. 
 
Our examination for this decay is similar to the $B$-violating decay 
in the preceding section. We search for the signal given by $t\bar t$ events
where one (say $t$) decays via  $L$-violating coupling,
$t\to  \mu^+  b   \tilde \chi^0_1$, while the other ($\bar t$) 
has the SM decays, $\bar t\to W^- \bar b$. 
Then there are two possible observing channels for such an
event: dilepton+2-jets and single lepton+4-jets, all being
associated  with missing energies. 
The dilepton channel has the lower rate and it is difficult to find a 
mechanism to enhance the S/B rate so as to find the "smoking gun" for 
the signal. So we search for the single lepton+4-jets channel which has
a higher rate. As is shown below, we can find effective selection cuts to 
enhance the S/B ratio for this signal.

Among the four jets in our signal there are two b-jets
(one is $b$, the other is $\bar b$). We require that at least
one b-jet passes b-tagging. The tagging efficiency is 
53\% at Run 1 and expected to reach 85\% at Run 2 and Run 3 
\cite{tev2000}. For the LHC we assume the tagging
efficiency to be the same as the Tevatron Run 2.  

So the signature is  $\ell+4j/b +\ETslash$ where $4j/b$ represents a 
4-jets event with at least one of the jets passing the $b$-tagging 
criterion.  This is the same as one of the typical signatures for 
$t\bar t$ event
in the SM, except for the different source of missing energy. 
To suppress the QCD background, we apply the basic selection cuts in 
Eqs.(\ref{basic1}-\ref{basic3}). Under the basic selection cuts the QCD 
background is
reduced to about 1/12 of the SM $t\bar t$ events~\cite{tev2000}.  
However, under the basic selection cuts the number of SM $t\bar t$ events 
far surpasses the number of signal events. In order to extract the signal
events, we turn to the transverse mass defined in Eq.(\ref{m_T}). 
For the SM $t\bar t$ events and $W$+jets background events 
the missing energy comes from the neutrino of the W decay, while for 
the signal
events the missing energy comes from the neutralino in the decay 
$t\to  \mu^+  \tilde b \to \mu^+ b   \tilde \chi^0_1$. Thus 
the transverse mass distributions of the SM background and 
the signal events are different, as shown in Fig.5.
In order to enhance the S/B ratio, we apply the following
cut, taking into account of the smearing effect,
\begin{equation}
m_{T}(\ell,p_T^{\rm miss}) \not\in 50\sim 100 {\rm~GeV}.
\end{equation}

Other details in the numerical calculation, such as the 
smearing of the energy of the final state particles and
the choice of SUSY parameters, are the same as in 
Sec.~\ref{sec2}.   
In Table 2 we present the signal cross section for sbottom mass 
of 150~GeV, with the comparison to the SM $t\bar t$ background. 
One sees that the transverse mass cut can enhance the S/B ratio
significantly. With the increase of sbottom mass, the signal cross
section drops rapidly, as shown in Table 3.

From Tables 2 and 3 one sees that Run 1 (0.1 fb$^{-1}$) of the Tevatron 
collider is unable to detect such decays for a sbottom heavier than 150~GeV 
and $\lambda'_{233}<1$. The possibility of observing such a decay is 
enhanced at Run 2 (2 fb$^{-1}$), Run 3 (30 fb$^{-1}$) and the LHC.
Under the discovery criteria $ S \ge 5 \sqrt{B}$,  the discovery limits 
of $\lambda'_{233}$ versus sbottom mass are plotted in Fig.6. The 
nonobservation of a signal is translated to the bounds (at 95\% C.L.)
shown in Fig.7.      

Since the current bounds on  $\lambda'_{233}$ from the LEP I Z-decay
are of ${\cal O}(1)$ for sfermion mass heavier than 100~GeV~\cite{Zdecay}, 
the results in Figs.6 and 7 indicate that the future runs at the upgraded 
Tevatron and LHC could either reveal the exotic decay or set stronger 
constraints on the $L$-violating coupling $\lambda'_{233}$. 

Our results for $\lambda'_{233}$ can be applied to the case of    
$\lambda'_{133}$. But for $\lambda'_{133}$ the current bound
from the $\nu_e$-mass, i.e.,  $\lambda'_{133}<0.0007$ at the $1-\sigma$
level~\cite{nue-mass}, is too strong, which makes the corresponding
decay $t\to  e^+  b   \tilde \chi^0_1$ unobservable.
 
\section{Summary and discussion}
\label{sec4}

We have examined the potential for the detection of top quark decays
via R-violating SUSY interactions at the Fermilab Tevatron and LHC.  
We studied two representative decay processes: one is
induced by the $B$-violating coupling $\lambda''_{331}$ and the other
is induced by the $L$-violating coupling $\lambda'_{233}$.
Both of them have a $b$-jet in their decay products and can proceed 
through the intermediate sbottom which was assumed to be light.
For the $B$-violating decay we searched for the signal 
$\ell+3j/2b +\ETslash$ given by $t\bar t$ events, while for the  
$L$-violating decay we searched for  the channel $\ell+4j/b +\ETslash$.  
We considered the possible backgrounds and performed a Monte Carlo simulation
by applying suitable cuts.
 
The signal cross section is found to drop drastically with
the increase of the intermediate sbottom mass. 
If the sbottom could be as light as $\sim 160$~GeV, then under the
current bounds of the relevant R-violating couplings, these decays
can be detectable at the future runs of the 
Tevatron and LHC. However, because of the small statistics, Run 1 of the 
Tevatron will not be adequate. 

A few remarks are due regarding our results:
\begin{itemize}
\item[{\rm(1)}] The results are sensitive to the sbottom mass; the signal is
                observable only for a light sbottom. The possibility of a
                light sbottom is usually motivated as follows:
                Firstly, the neutral kaon system gives a strong  
                constraint~\cite{KKbar} on the masses of the first and second 
                generation squarks. The third generation sfermions are much 
		less constrained so far.  Secondly, in the supergravity 
		scenario of supersymmetry breaking, mass splitting of the 
		third-generation and the other sfermions results from 
		the renormalization group evolution of the masses between the 
		unification scale and the weak scale, even if the sfermions 
		have equal masses at the unification scale.  This splitting 
		is due to the effect of the large Yukawa coupling of the top. 
		The bottom and tau sectors are also affected. Thirdly, there 
		are arguments~\cite{heavysf} that first and second generation 
		sfermions can be as heavy as 10 TeV without conflicting the 
		naturalness problem, while the third generation sfermions have
		to be rather light.

\item[{\rm(2)}] As pointed out in Sec.~(\ref{sec1}), the two decay  
		processes we considered resemble the favorable cases in which
		a $b$-jet is produced in the decay products. 
		While we can apply our results directly to the cases of 
		$\lambda''_{332}$ and $\lambda'_{133}$, we noticed that 
		similar decays induced by other couplings like 
	      $\lambda''_{312}$ and $\lambda'_{232}$ give poor signals 
	      since there is no $b$-quark in their corresponding top decays.   

\item[{\rm(3)}] We noted that apart from the relevant  R-violating couplings 
 		and the sbottom mass, our results are also dependent on the 
		mass and coupling of the lightest neutralino. In our 
		calculation we only present some illustrative results by 
		fixing a set of SUSY parameters rather than scanning 
		the entire allowed SUSY parameter space. In some unfavorable
		cases, such as when the mass of the lightest neutralino (LSP) 
                is 
		close to the sbottom mass so that the b-quark from the sbottom 
		decay ($ \tilde b \to b   \tilde \chi^0_1$) is too soft to pass
		the selection cuts, these exotic decays would be unobservable 
		even at the LHC.

\item[{\rm(4)}] As pointed out in Sec. \ref{sec2}, the $B$-violating decay 
                gives the unique signal of same sign b-quarks while the
                main SM backgrounds give the unlike sign b-quarks. 
                To be conservative, we assumed in our analysis that the $b$ 
                tagging is not of sufficient sensitivity to distinguish 
                between a $b$-jet and a $\bar b$-jet. If $b$ charge 
                identification can be achieved in future 
                detectors, more stringent discovery limits than
                those we have presented will be possible.
                Additional improvements will be possible if hadronic
                jets from $\tau$ decays can be clearly identified as 
                such, thus reducing the background from $\tau$ 
                hadronization.
\end{itemize}

\section*{ Acknowledgments }
We thank E. Boos and L. Dudko for the discussion of  $Wb\bar b j$ 
background.
BLY acknowledges the hospitality extended to him by Professor Zhongyuan Zhu
and colleagues at the Institute of Theoretical Physics, Academia Sinica, 
where part of the work was performed. 
This work is supported in part by a grant of Chinese Academy of Science
for Outstanding Young Scholars and also by a DOE grant No. DE-FG02-92ER40730.


\newpage
\null\vspace{0.4cm}
\noindent
{\small Table 1: Signal $\ell+3j/2b +\ETslash$ and  background cross sections 
                 in units of fb. The basic cuts are 
                 $p_T^{\rm all}\ge 20 \rm{~GeV}$, 
                 $\vert\eta_{\rm all}\vert \le 2.5$ and $\Delta R \ge 0.5$, 
                 and the transverse mass cut is $m_T\ge 120$~GeV.
                 The signal results were calculated by 
                 assuming $M=100$~GeV, $\mu=-200$~GeV 
                 and $\tan\beta=1$. The double b-jet tagging with $42 \%$ 
                 efficiency is assumed. The charge conjugate channels have 
                 been included.}
\vspace{0.2cm}
\begin{center}
\begin{tabular}{|c|lccccccc|}
\hline
 &Sbottom mass(GeV)         & 150 & 155 & 160 & 165 & 170 & 180 & 190\\
Tevatron   
&Signal/$(\lambda''_{331})^2$ & 11 & 5.8 & 2.04& 0.27& 0.01& 0.005&0.003\\
 (1.8 Tev)
 &Background               & 2.07 & & & & & &  \\ \hline

 &Sbottom mass(GeV)           & 150 & 155 & 160 & 165 & 170 & 180 & 190\\
Tevatron            
&Signal/$(\lambda''_{331})^2$& 16  & 8.4 & 3.0 & 0.4 & 0.02&0.007&0.004\\ 
 (2 Tev) 
&Background               & 3.05  & & & & & &   \\ \hline

 &Sbottom mass(GeV)         & 150 & 155 & 160 & 165 & 170 & 180 & 190\\
LHC    		   
&Signal/$(\lambda''_{331})^2$ & 1624& 885 & 371 & 58  & 1.7 & 0.4 &  0.3\\
 (14 Tev) 
 &Background               & 350 & & & & & &  \\ \hline
\end{tabular}
\end{center}
\vspace{1cm}

\newpage
\null\vspace{0.4cm}
\noindent
{\small Table 2: Signal $\ell+4j/2b +\ETslash$ and  the SM $t\bar t$ 
                 background cross sections for sbottom mass of 150~GeV. 
                 The basic cuts are 
                 $p_T^{\rm all}\ge 20 \rm{~GeV}$, 
                 $\vert\eta_{\rm all}\vert \le 2.5$ and $\Delta R \ge 0.5$, 
                 and the transverse mass cut is 
                 $m_{T}(\ell,p_T^{\rm miss}) \not\in 50\sim 100 {\rm~GeV}$.
                 The signal results were calculated by 
                 assuming $M=100$~GeV, $\mu=-200$~GeV 
                 and $\tan\beta=1$. Tagging at least one $b$-jet is assumed 
                 for 53\% efficiency for the Tevatron (1.8 TeV), $85 \%$ 
                 efficiency for the upgraded Tevatron (2 TeV) and LHC.
                 The charge conjugate channels have been included.}
\vspace{0.1in}
\begin{center}
\begin{tabular}{|l|l|c|c|} \hline
     & &            &                              \\ 
     & & basic cuts & basic cuts                    \\
     & &            & $+$                           \\ 
     & &            &$m_T(\ell,p_T^{\rm miss})$ cut \\  
                                                      \hline  
     & Signal/$(\lambda'_{233})^2$ (fb)  &70  & 43    \\ \cline{2-4}
 Tevatron (1.8 TeV) 
     & Background (fb)                   &300 & 86     \\
                                                         \hline
     & Signal/$(\lambda'_{233})^2$ (fb)  &154 & 96    \\ \cline{2-4}
 Tevatron (2 TeV) 
     & Background (fb)                   &662 & 193   \\
                                                         \hline
     & Signal/$(\lambda'_{233})^2$ (pb)  &12.7& 8.2    \\ \cline{2-4}
 LHC (14 TeV) 
     & Background (pb)                   &54  & 16     \\
                                                         \hline

\end{tabular}
\end{center}
\vspace{1cm}

\newpage
\null\vspace{0.4cm}
\noindent
{\small Table 3: Same as Table 2, but for the signal cross section
                 versus sbottom mass
                 under the basic plus transverse mass cut.}
\vspace{0.2cm}
\begin{center}
\begin{tabular}{|l|cccccccc|}
\hline
Tevatron (1.8 Tev): & &  & &  & &  &  &\\ 
 Sbottom mass(GeV)         & 150 & 155 & 160 & 165 & 170 & 180 & 190 &200\\
Signal/$(\lambda'_{331})^2$ (fb)
 & 42.8& 23.7 &8.0 & 0.86 &0.04 &0.02&  0.01& 0.007
                               \\ \hline
Tevatron (2 TeV):  & &  & &  & &  &  &\\ 
 Sbottom mass(GeV)           & 150 & 155 & 160 & 165 & 170 & 180 & 190&200\\
 Signal/$(\lambda'_{233})^2$ (fb)
  &  96 &  53 &  19 &  2.2 & 0.09 &0.04 & 0.024&
                                0.016
\\ \hline

LHC (14 TeV): & &  & &  & &  &  &\\    		   
Sbottom mass(GeV)         & 150 & 155 & 160 & 165 & 170 & 180 & 190&200\\
Signal/$(\lambda''_{233})^2$ (pb)
 & 8.2 &  4.8&  1.86& 0.26& 0.008 &0.003 &0.002& 0.001 \\ \hline
\end{tabular}
\end{center}
\vspace{1cm}

\newpage
\begin{figure}[htb]
\vspace*{-3cm}
\hspace*{-4cm}
 \includegraphics[height=27cm,width=22cm, angle=0]{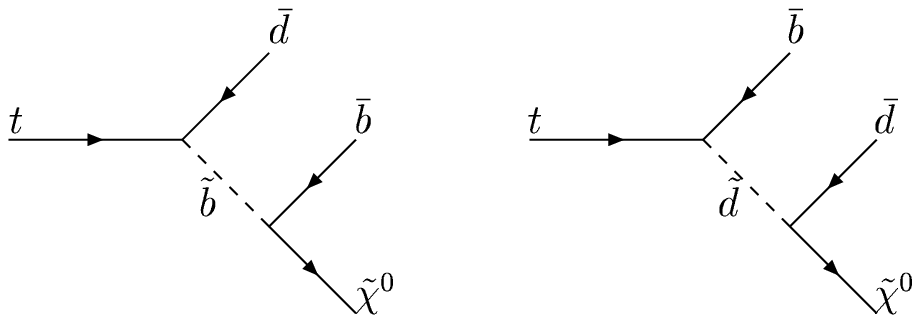}
\vspace*{-20cm}
\caption[]{ The Feynman diagram for the $B$-violating decay 
            induced by $\lambda''_{331}$. }
\label{fig1}
\end{figure}
\begin{figure}[htb]
\hspace*{-4cm}
 \includegraphics[height=27cm,width=22cm, angle=0]{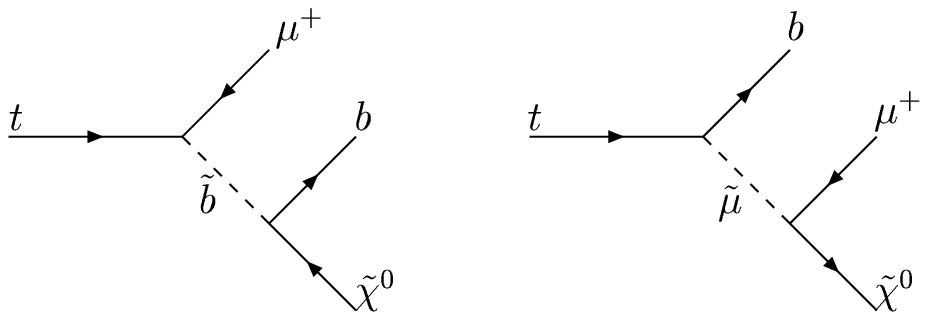}
\vspace*{-20cm}
\caption[]{ The Feynman diagram for the $L$-violating decay 
            induced by $\lambda'_{233}$. }
\label{fig2}
\end{figure}

\newpage
\begin{figure}[htb]
 \includegraphics[height=15cm,width=12cm, angle=90]{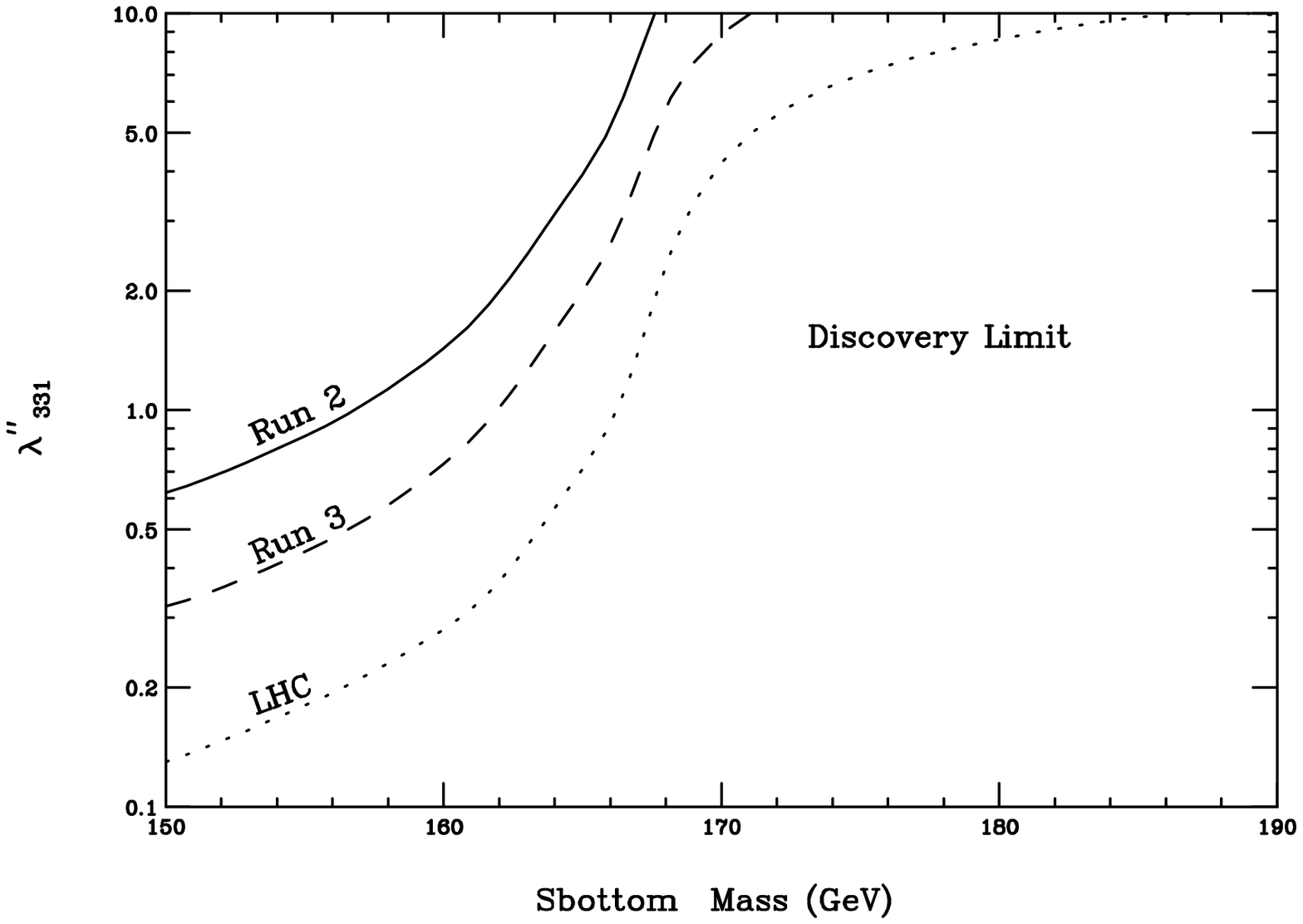}
\caption[]{ The discovery limits of $\lambda''_{33j}$ versus 
            sbottom mass at Run 2 (2 fb$^{-1}$), Run 3 (30 fb$^{-1}$)
            and LHC (10 fb$^{-1}$). The region above each curve is the
            corresponding region of discovery.} 
\label{fig3}
\end{figure}

\newpage
\begin{figure}[htb]
 \includegraphics[height=15cm,width=12cm, angle=90]{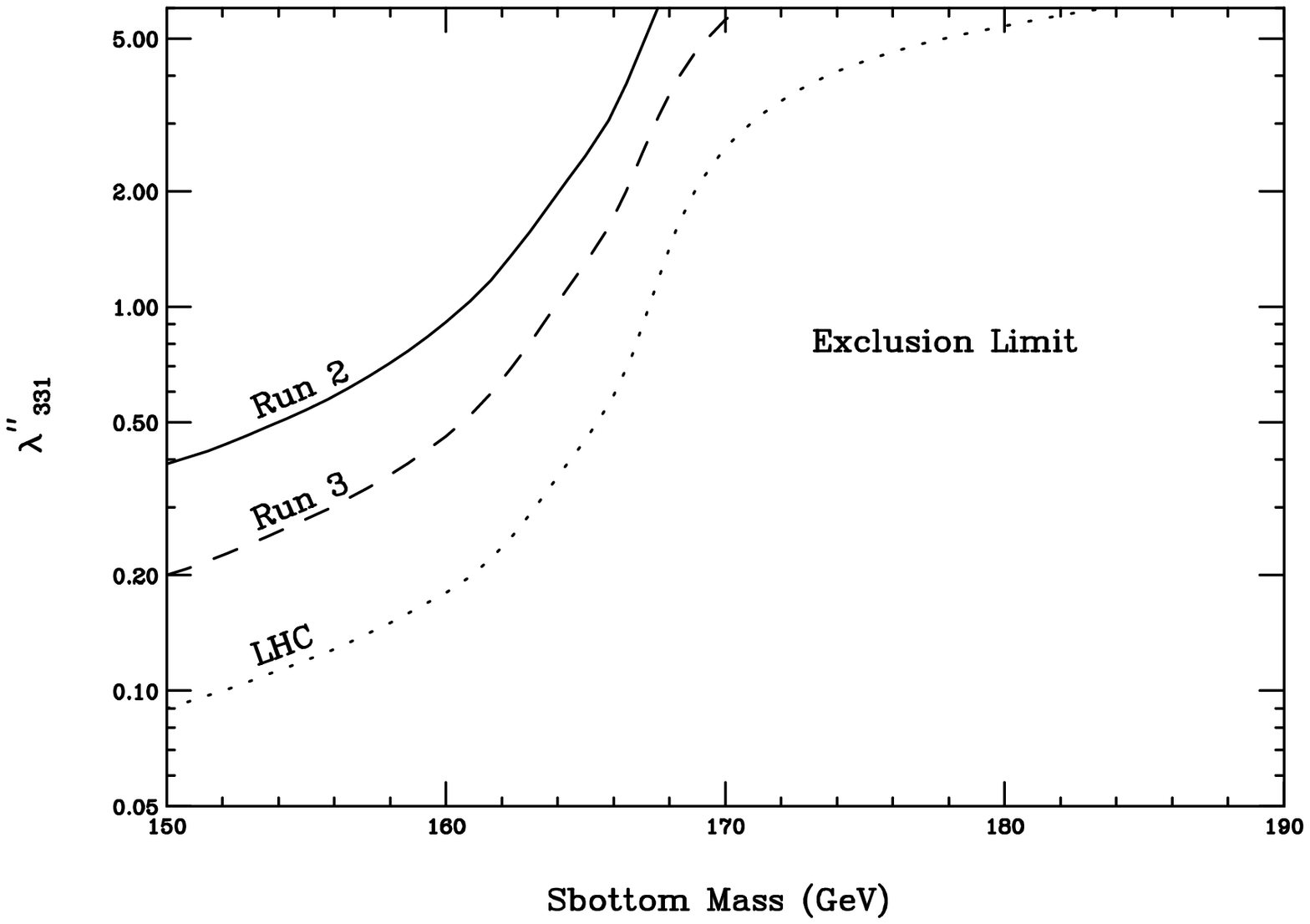}
\caption[]{ The exclusion limits of $\lambda''_{33j}$ versus 
            sbottom mass at Run 2 (2 fb$^{-1}$), Run 3 (30 fb$^{-1}$)
            and LHC (10 fb$^{-1}$). The region above each curve is the
            corresponding region of exclusion.} 
\label{fig4}
\end{figure}

\newpage
\begin{figure}[htb]
 \includegraphics[height=15cm,width=12cm, angle=90]{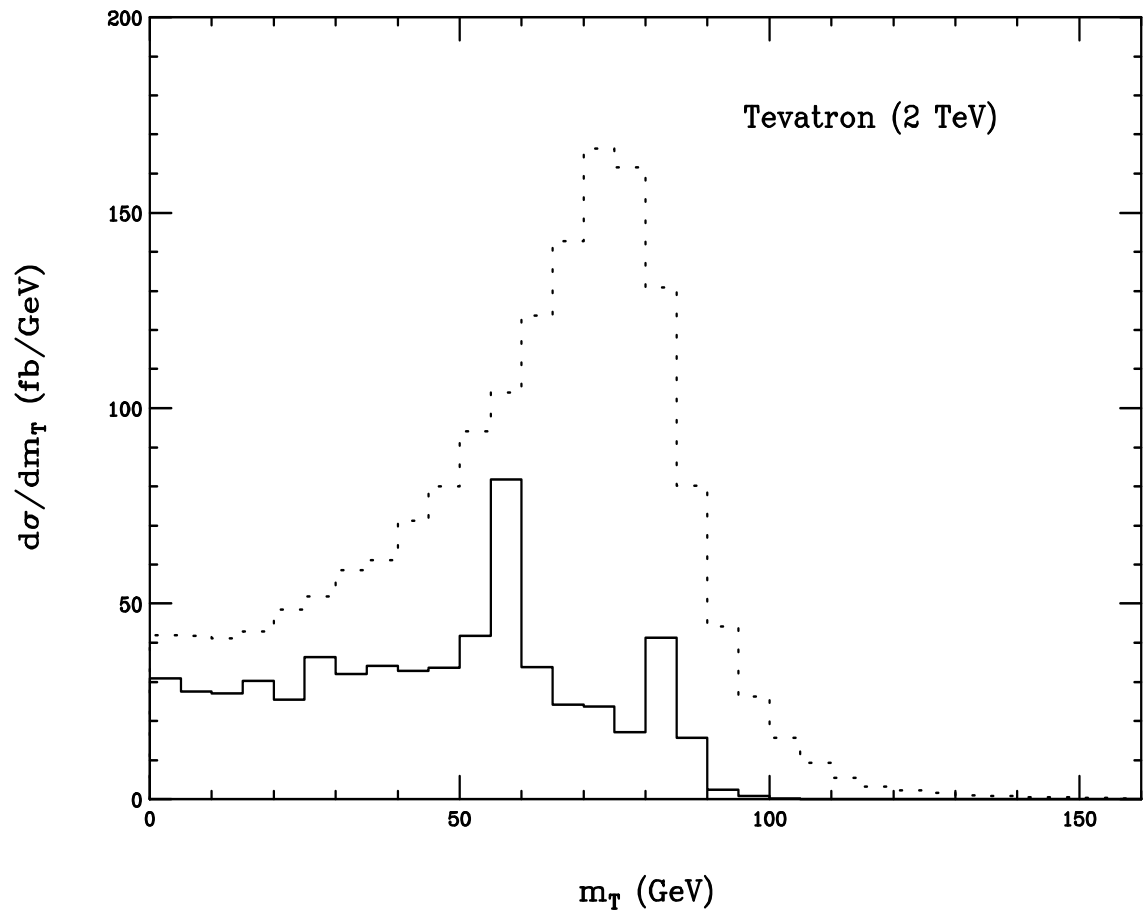}
\caption[]{ The transverse mass, $m_T(\ell,p_T^{\rm miss})$,
distribution of $\ell+4j/b+\ETslash$ at the Tevatron collider.
The solid curve is for the signal event.  The dotted curve is for the 
SM $t\bar t$ background.}
\label{fig5}
\end{figure}

\begin{figure}[htb]
 \includegraphics[height=15cm,width=12cm, angle=90]{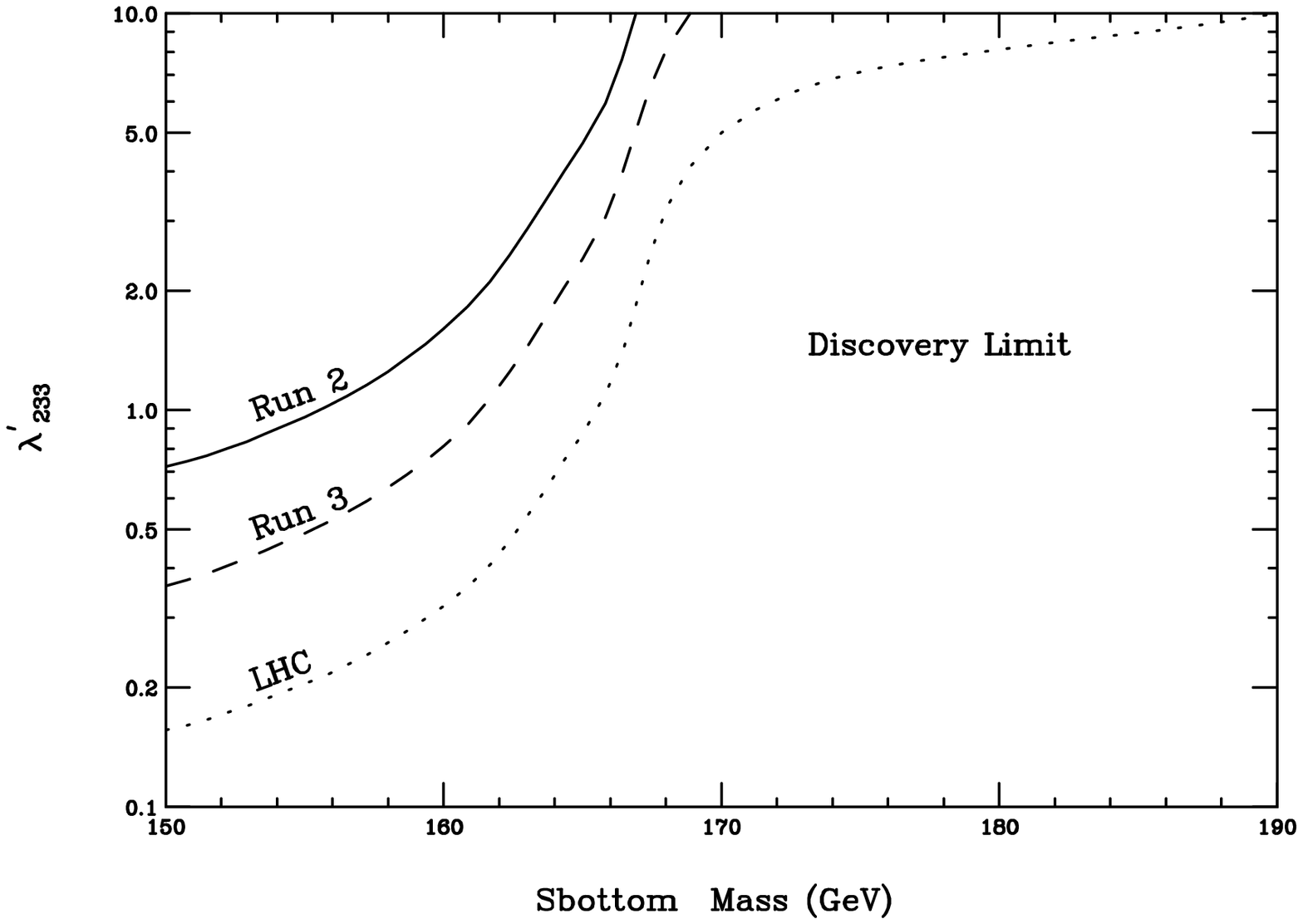}
\caption[]{ The discovery limits of $\lambda'_{233}$ versus 
            sbottom mass at Run 2 (2 fb$^{-1}$), Run 3 (30 fb$^{-1}$)
            and LHC (10 fb$^{-1}$). The region above each curve is the
            corresponding region of discovery.} 
\label{fig6}
\end{figure}

\newpage
\begin{figure}[htb]
 \includegraphics[height=15cm,width=12cm, angle=90]{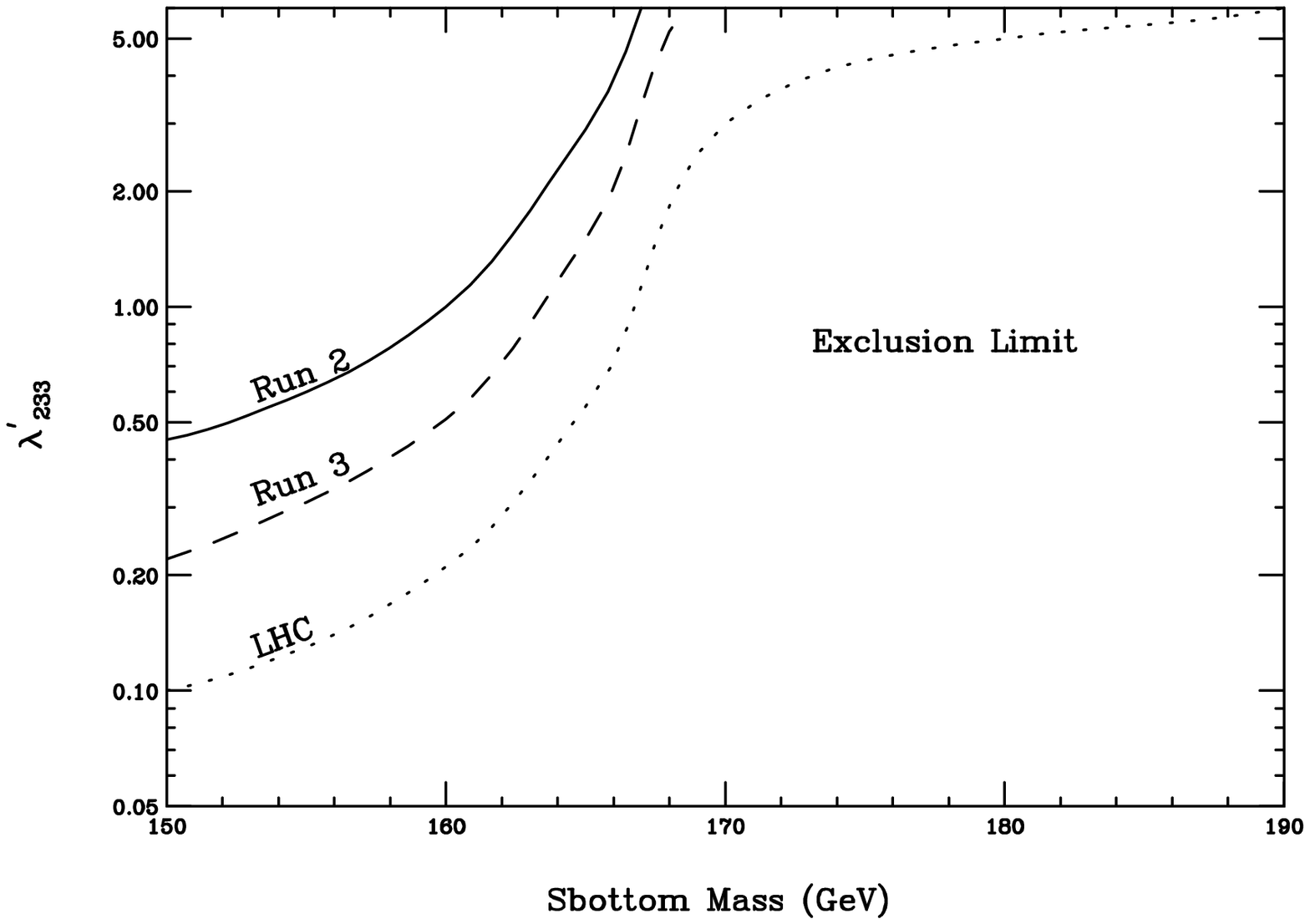}
\caption[]{ The exclusion (95\% C.L.) limits of $\lambda'_{233}$ versus 
            sbottom mass at Run 2 (2 fb$^{-1}$), Run 3 (30 fb$^{-1}$)
            and LHC (10 fb$^{-1}$). The region above each curve is the
            corresponding region of exclusion.} 
\label{fig7}
\end{figure}
\end{document}